\documentclass[aps,prb,twocolumn,showpacs,amsmath,amssymb]{revtex4}
\usepackage{graphicx}
\usepackage{bm}

\begin{document}

\title{Tunneling conductance of ferromagnet/noncentrosymmetric superconductor junctions}

\author{S. Wu and K. V. Samokhin}

\affiliation{Department of Physics, Brock University, St. Catharines, Ontario L2S 3A1, Canada}

\date{\today}

\begin{abstract}
Based on the extended Blonder-Tinkham-Klapwijk formalism, the
tunneling conductance characteristics of a planar junction between
a ferromagnet and a noncentrosymmetric superconductor are studied.
The effects of the Rashba spin-orbit coupling (RSOC), the exchange
energy, and the Fermi wave-vector mismatch (FWM) on the conductance are
all taken into account. In the absence of the FWM, it is found that far
away from the gap edge the conductance is suppressed by the
RSOC, while around the gap edge it is almost independent of RSOC. The interplay of the RSOC and the exchange energy causes an
enhancement of the subgap conductance, which is more pronounced
when the RSOC is small. When the FWM is introduced, it is shown that the
conductance is monotonically enhanced as the FWM parameter decreases.
\end{abstract}


\maketitle

\section{Introduction}

In recent years, tunneling spectroscopy has played a crucial role in
probing electronic states of superconductors. In normal
metal/superconductor (N/S) junctions, zero-bias conductance peaks
(ZBCP)\cite{prl9472-1526,prb9653-2667,prb9756-13746} observed in
high-temperature superconductors (HTSC) are explained as arising
from the sign change of the pair potential, which leads to the
formation of midgap surface states. Replacing the normal metal by a
ferromagnetic metal, the conductance spectrum is considerably changed due to the
spin polarization caused by the exchange field.
Earlier works\cite{prl9574-1657,science98282-85,prb0367-020411,prl9881-3247,prb9859-9558}
have demonstrated that the effect of the exchange energy is,
in general, to reduce the Andreev reflection (AR) at a ferromagnet/centrosymmetric superconductor (FM/CSC) interface. 
So far, a variety of physical phenomena, including the effects of
temperature,\cite{prb0367-174501} the planar magnetization
components,\cite{prb0775-134509} and the FWM
\cite{prb9960-6320,prb0061-1555} on the tunneling conductance and the
proximity effect\cite{prb0265-174508,jap08104-063903} have been
investigated. In particular, in Refs. \onlinecite{prb9960-6320} and
\onlinecite{prb0061-1555} the effect of the FWM was considered and it was found
that in some cases the exchange energy can enhance Andreev
reflection.

The recent discovery of superconductivity in the heavy fermion
compound CePt$_{3}$Si (Ref. \onlinecite{prl0492-027003}) has renewed interest,
both experimental and theoretical, in the properties of
superconductors without inversion symmetry. Noncentrosymmetric
superconductors (NCSCs) exhibit a variety of distinctive features, which are
absent in the centrosymmetric case, such as a strongly anisotropic
spin susceptibility with a large residual
component,\cite{prl0187-037004,prb0265-144508,prl0594-027004}
magnetoelectric effect,\cite{prl9575-2004,prb0572-024515} and unusual
nonuniform (``helical'') superconducting
phases.\cite{pc03387-13,prb0470-104521,prl0594-137002} The
tunneling conductance in a normal
metal/noncentrosymmetric superconductor (N/NCSC) junction has been
recently studied in Refs. \onlinecite{prb0572-220504,prb0776-012501,prb0776-054511}.
In these works, Yokoyama {\it et al.}\cite{prb0572-220504} found
that an intrinsically $s$-wave-like property of a triplet NCSC
results in a peak at the energy gap in the
tunneling spectrum. Iniotakis {\it et al.}\cite{prb0776-012501}
observed the zero-bias anomalies if a specific form of the mixed
singlet-triplet order parameter was realized. Linder {\it et al.}
\cite{prb0776-054511} found pronounced peaks and bumps in the
conductance spectrum corresponding to the sum and difference of the
magnitudes of the singlet and triplet gaps. One of the important
questions is how the Andreev reflection affects the tunneling conductance in the
presence of \textit{both} ferromagnetism and the RSOC. So far, there has been no theory
for this phenomenon.

The purpose of this paper is to investigate the
tunneling spectroscopy of a ferromagnet/noncentrosymmetric
superconductor (FM/NCSC) junction. We employ the well-known
Blonder-Tinkham-Klapwijk (BTK) formalism,\cite{prb8225-4515} but
extend and generalize it to include the effects of the exchange
energy (some references called it spin polarization) in the
ferromagnet, the RSOC due to the lack of inversion symmetry,
and the existence of FWM. We find many interesting
features in the conductance spectrum, stemming from the interplay of
magnetism and the RSOC. Away from the gap edge, the
tunneling conductance is enhanced as the RSOC decreases, while it is
almost unchanged near the gap edge. This behavior is
completely different from that found in the N/NCSC
junction.\cite{prb0572-220504} The competition between the effects of
the exchange energy and the RSOC on the AR leads to an enhanced
subgap conductance, which can even result in a maximum at zero energy under certain
conditions. In addition, we also show the importance of properly
accounting for the FWM, namely, the conductance spectrum monotonically
increase with the decreasing the FWM parameter in the whole excitation energy
region, which is essentially different from the behaviour found in the FM/CSC
junctions.\cite{prb9960-6320,prb0061-1555,prb0979-014502}

The paper is organized as follows: In Sec.~II, we define the theoretical model and extend the BTK
approach to obtain the amplitudes for various scattering processes
that occur in the FM/NCSC junction. In Sec.~III, the corresponding
numerical results for the tunneling conductance are presented and discussed.
Sec.~IV contains a summary of our results.

\section{Formulation of the model}

\begin{figure}
\includegraphics[width=9cm]{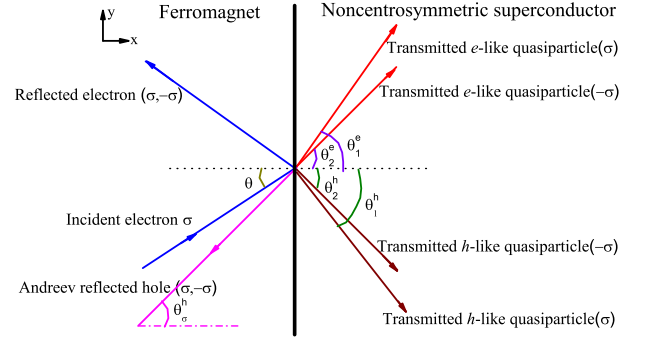}
\caption{(Color online) Schematic illustration of the scattering
processes at the FM/NCSC interface. The angles of normal and Andreev reflection for electrons and holes with $\sigma=\uparrow,\downarrow$ are different.
Due to the presence of spin-orbit coupling, the electron-like and hole-like excitations on the
superconducting side are scattered through different angles.}
\label{fig: scattering}
\end{figure}

We consider the tunneling conductance of the FM/NCSC junction as
shown in Fig. \ref{fig: scattering}. The FM is at $x<0$, and is described by an
effective single-particle Hamiltonian. The NCSC is assumed to have
purely singlet pairing, and is described by a BCS-like Hamiltonian.
The FM/NCSC interface is at $x=0$, where there is interfacial
scattering, which is modeled by a potential $U(\bm{r})=U_{\textrm{0}}\delta(x)$, with $U_{\textrm{0}}$ characterizing the barrier strength. 
The band dispersions are isotropic, and the effective masses of quasiparticles are
assumed to be the same on both sides. The quasiparticle
wave function satisfies the following Bogoliubov-de Gennes (BdG)
equation:
\begin{equation}
	{\cal H}\Psi(\bm{r})=E\Psi(\bm{r}),
\end{equation}
where
\begin{equation}
	{\cal H}=\left(\begin{array}{cc}
	\hat{H}(\bm{r})-\sigma h(\bm{r}) & \hat{\Delta}(\bm{r})\\
	\hat{\Delta}^{\dag}(\bm{r}) & -(\hat{H}^{T}(\bm{r})+\sigma
	h(\bm{r}))
	\end{array}\right),
\end{equation}
with the single-particle Hamiltonian 
$$
	\hat{H}(\bm{r})=\left(-\frac{\bm{\nabla}^{2}}{2m}+U(\bm{r})-E_{Fi}\right)\hat\sigma_0+{\bm\gamma}({\bm k},{\bm r}) \bm{\hat{\sigma}}.
$$
Here $E_{Fi}=E_{FM},E_{FS}$ represent the Fermi energies in the FM and
the NCSC region, respectively, 
$\sigma=\pm 1$ for different spin orientations,
$h(\bm{r})=h_{\textrm{0}}\theta(-x)$ is the exchange energy on the
FM side (we assume that the FM magnetization and the exchange energy are along the $z$ axis), $\bm{\gamma}(\bm{k},\bm{r})=\bm{\gamma}(\bm{k})\theta(x)$ is the antisymmetric (Rashba) spin-orbit coupling on the SC side,
and $\bm{\hat{\sigma}}$ are the Pauli matrices (we use the units in which $\hbar=1$). 

In our model, we consider a
noncentrosymmetric superconductor with the tetragonal crystal
symmetry, which is relevant for CePt$_{3}$Si, CeRhSi$_{3}$, and
CeIrSi$_{3}$. We choose the RSOC in the following form:
$\bm{\gamma}(\bm{k})=\gamma_{\textrm{0}}(k_{y},-k_{x},0)$ with the 
Rashba coupling constant $\gamma_{\textrm{0}}$ and the BCS pairing
potential
$\hat{\Delta}(\bm{r})=i\hat\sigma_{y}\Delta_{\textrm{0}}\theta(x)$. We
take into account the fact that the Fermi energy to be different in the FM and
NCSC regions, which allows for different bandwidths originating
from different carrier densities in the two regions. We introduce the dimensionless FWM parameter as follows:
$R=k_{FS}/k_{FM}\equiv\sqrt{E_{FS}/E_{FM}}$. In the next section we will show that the FWM between the two
regions plays an important role in the tunneling conductance.

We focus on the excitations with $E\geq0$, assuming an incident
electron above the Fermi level. When an electron is injected from the FM side, with spin $\sigma=\uparrow,\downarrow$, the excitation
energy $E$, and the wave vector $\bm{k}^{e}_{\sigma}$, at an angle
$\theta$ from the interface normal, there are four reflection processes: (i)
Andreev reflection to the majority spin ($r^{\uparrow}_{h}$), (ii)
Andreev reflection to the minority spin ($r^{\downarrow}_{h}$), (iii)
normal reflection to the majority spin ($r^{\uparrow}_{e}$), (iv) normal
reflection to the minority spin ($r^{\downarrow}_{e}$), see Fig.~1. The Andreev and
normal reflection coefficients are denoted by $r^{\sigma}_{h}$ and
$r^{\sigma}_{e}$, respectively. Solving the BdG equation, the wave
function is $\Psi(\bm{r})=\Psi(x)e^{i{\bm k_{\parallel}}{\bm
r_{\parallel}}}$, where $\bm{r}_\parallel$ is parallel to the interface, and
\begin{eqnarray}
	\Psi_{FM}(x)=\left(\begin{array}{c}
	s\\ 0\\ 0\\0
	\end{array}\right)e^{ik_{\uparrow}^{e}\cos\theta x}
	+\left(\begin{array}{c}
	0\\ \bar{s}\\ 0\\ 0
	\end{array}\right)e^{ik_{\downarrow}^{e}\cos\theta x}
	\quad\nonumber\\
	+r_{e}^{\uparrow}
	\left(\begin{array}{c}
	1\\ 0\\ 0\\ 0
	\end{array}\right)e^{-ik_{\uparrow}^{e}Ax}
	+r_{e}^{\downarrow}
	\left(\begin{array}{c}
	0\\ 1\\ 0\\ 0
	\end{array}\right)e^{-ik_{\downarrow}^{e}\Bar{A}x}\nonumber\\
	+r_{h}^{\uparrow}
	\left(\begin{array}{c}
	0\\ 0\\ 1\\ 0
	\end{array}\right)e^{ik_{\uparrow}^{h}\cos\theta^{h}_{\uparrow}x}
	+r_{h}^{\downarrow}
	\left(\begin{array}{c}
	0\\ 0\\ 0\\ 1
	\end{array}\right)e^{ik_{\downarrow}^{h}\cos\theta^{h}_{\downarrow}x}
\end{eqnarray}
on the FM side. The notations are as follows: The quasiparticle wave
vectors are given by
\begin{eqnarray*}
	&& k_{\uparrow}^{e(h)}=\sqrt{2m[E_{FM}+(-)E+h_{\textrm{0}}]},\\
	&& k_{\downarrow}^{e(h)}=\sqrt{2m[E_{FM}+(-)E-h_{\textrm{0}}]}.
\end{eqnarray*}
An incoming electron with spin-$\uparrow$ is described by $s=1,\bar{s}=0$, while a spin-$\downarrow$ electron by
$s=0,\bar{s}=1$. Then, $A=s\cos\theta+\bar{s}\cos\theta_{\uparrow}^{e}$ and
$\bar{A}=\bar{s}\cos\theta+s\cos\theta_{\downarrow}^{e}$, and $\theta_{\sigma}^{e(h)}$ are angles between the wave vectors
$\bm{k}_{\sigma}^{e(h)}$ and the interface normal.

Similarly, the BdG wave function on the superconducting side is given by
\begin{eqnarray}
	\Psi_{SC}(x)=\frac{t_{e}^{\uparrow}}{\sqrt{2}}
	\left(\begin{array}{c}
	u\\ -ie^{i\theta^{e}_{1}}u\\ ie^{i\theta^{e}_{1}}v\\ v
	\end{array}\right)e^{ik_{1}^{e}\cos\theta_{1}^{e}x}
	\quad\quad\nonumber\\
	+\frac{t_{e}^{\downarrow}}{\sqrt{2}}
	\left(\begin{array}{c}
	u\\ ie^{i\theta^{e}_{2}}u\\ -ie^{i\theta^{e}_{2}}v\\ v
	\end{array}\right)e^{ik_{2}^{e}\cos\theta_{2}^{e} x}
	\quad\quad\nonumber\\
	+\frac{t_{h}^{\uparrow}}{\sqrt{2}}
	\left(\begin{array}{c}
	v\\ ie^{-i\theta^{h}_{1}}v\\ -ie^{-i\theta^{h}_{1}}u\\ u
	\end{array}\right)e^{-ik_{1}^{h}\cos\theta_{1}^{h} x}
	\quad\quad\nonumber\\
	+\frac{t_{h}^{\downarrow}}{\sqrt{2}}
	\left(\begin{array}{c}
	v\\ -ie^{-i\theta^{h}_{2}}v\\ ie^{-i\theta^{h}_{2}}u\\ u
	\end{array}\right)e^{-ik_{2}^{h}\cos\theta_{2}^{h} x},
\end{eqnarray}
with the wave vectors 
\begin{eqnarray*}
 	&& k_{1}^{e(h)}=-m\gamma_{\textrm{0}}+\sqrt{(m\gamma_{\textrm{0}})^{2}+2m[E_{FS}+(-)\Omega]},\\
	&& k_{2}^{e(h)}=m\gamma_{\textrm{0}}+\sqrt{(m\gamma_{\textrm{0}})^{2}+2m[E_{FS}+(-)\Omega]},
\end{eqnarray*}
and $\Omega=\sqrt{E^{2}-\Delta_{\textrm{0}}^{2}}$. The transmission amplitudes of
electron-like and hole-like quasiparticles are $t_{e}^{\sigma}$ and $t_{h}^{\sigma}$, respectively. 
The coherence factors in the NCSC region are given as
\begin{eqnarray}
	u=\frac{1}{\sqrt{2}}\sqrt{1+\frac{\Omega}{E}},\qquad v=\frac{1}{\sqrt{2}}\sqrt{1-\frac{\Omega}{E}}.
\end{eqnarray}
Finally, $\theta_{1(2)}^{e(h)}$ are the angles between the wave vectors
$k_{1(2)}^{e(h)}$ and the interface normal, as shown in Fig.~1. 
The angles are obtained from the following equations:
\begin{eqnarray}
	(sk_{\uparrow}^{e}+\bar{s}k_{\downarrow}^{e})\sin\theta=sk_{\downarrow}^{e}\sin\theta_{\downarrow}^{e}+
	\bar{s}k_{\uparrow}^{e}\sin\theta_{\uparrow}^{e}\quad\nonumber\\
	=k_{\sigma}
	^{h}\sin\theta_{\sigma}^{h}=k_{1(2)}^{e(h)}\sin\theta_{1(2)}^{e(h)},
\end{eqnarray}
which express the conservation of the parallel component of the wave
vector due to the translational symmetry along the interface. 

All the coefficients in Eqs. (3) and (4) can be determined by
the following boundary conditions for the wave functions:
\begin{equation}
	\Psi_{FM}|_{x=0^{-}}=\Psi_{SC}|_{x=0^{+}},
\end{equation}
\begin{equation}
	\hat v_{x}\Psi_{SC}|_{x=0^{+}}-\hat v_{x}\Psi_{FM}|_{x=0^{-}}=-2iU_0\eta\Psi_{FM}|_{x=0^{-}},
\end{equation}
where $\eta$ is the $4\times4$ matrix
\begin{equation}
	\eta=\left(\begin{array}{cccc}
	1 & 0 & 0 & 0\\
	0 & 1 & 0 & 0\\
	0 & 0 & -1 & 0\\
	0 & 0 & 0 & -1
	\end{array}\right),
\end{equation}
and the velocity operator in the $x$-direction is defined as\cite{prb0164-121202}
\begin{equation}
	\hat v_x=
	\left(\begin{array}{cccc}
	-\frac{i}{m}\frac{\partial}{\partial x} & i\gamma_{\textrm{0}}\theta(x) & 0 & 0\\
	-i\gamma_0\theta(x) & -\frac{i}{m}\frac{\partial}{\partial x} & 0 & 0\\
	0 & 0 & \frac{i}{m}\frac{\partial}{\partial x} & -i\gamma_0\theta(x)\\
	0 & 0 & i\gamma_0\theta(x) &
	\frac{i}{m}\frac{\partial}{\partial x}
	\end{array}\right).
\end{equation}
Note that the presence of the spin-orbit coupling results in the off-diagonal components of the velocity operator.
We also introduce the dimensionless parameters $Z=2mU_{\rm0}/k_{FS}$
and $\alpha=2m\gamma_{\rm 0}/k_{FS}$, characterizing the barrier
strength and the magnitude of the RSOC, respectively.

By using the general BTK formalism,\cite{prb8225-4515} we obtain for the
dimensionless differential tunneling conductance:
\begin{eqnarray}
	&& G(E)=\sum\limits_{\sigma}P_{\sigma}G_{\sigma}(E),\\
	&& G_{\sigma}(E)=\frac{1}{G_{N}}\int_{\theta_{c}}d\theta\cos\theta\, G_{\sigma}(E,\theta),\nonumber\\
	&& G_{N}=\int_{\theta_{c}}d\theta\cos\theta\frac{4\cos^{2}\theta}{4\cos^{2}\theta+Z^{2}},\nonumber
\end{eqnarray}
where $P_{\sigma}=\frac{1}{2}(1+\sigma h_{\rm 0}/E_{FM})$ is the
probability that an incident electron has spin $\sigma$ ($P_\uparrow\neq P_\downarrow$ because of the difference between the densities of states in the spin-$\uparrow$ and spin-$\downarrow$ bands, see Ref. \onlinecite{prl9574-1657}), $G_{N}$ is the tunneling conductance for a
normal metal/normal metal junction, and $\theta_{c}$ is determined by
the angle of total reflection (critical angle) for incident electron
with spin $\sigma$. For an incoming electron with spin-$\uparrow$,
the critical angles for the Andreev reflection and the transmission
are given by
$\theta_{c1}=\arcsin(k_{\downarrow}^{h}/k_{\uparrow}^{e})$ and
$\theta_{c2}=\arcsin(k_{1}^{e(h)}/k_{\uparrow}^{e})$, respectively.
When $\theta$ exceeds $\theta_{c1}$, the $x$-component of the wave
vector in the AR process,
$\sqrt{(k_{\downarrow}^{h})^{2}-(k_{\uparrow}^{e})^{2}\sin^{2}\theta}$,
becomes purely imaginary so that the Andreev-reflected
quasiparticles do not contribute to the charge current, which can be
referred to as the virtual AR. Further, when $\theta>\theta_{c2}$, the
transmitted quasiparticles with the wave vectors $k_{1}^{e(h)}$ do not contribute to the conductance. 

The conductance for an electron with spin $\sigma$ as a function of the excitation energy $E$ and the incident angle $\theta$ reads
\begin{equation}
	G_{\sigma}(E,\theta)=1+\frac{\lambda_{1}}{\lambda_{0}}|r_{h}^{\uparrow}|^{2}
	+\frac{\lambda_{2}}{\lambda_{0}}|r_{h}^{\downarrow}|^{2}
	-\frac{\lambda_{3}}{\lambda_{0}}|r_{e}^{\uparrow}|^{2}-\frac{\lambda_{4}}{\lambda_{0}}|r_{e}^{\downarrow}|^{2}.
\end{equation}
The ratios of $\lambda_{i}$ on the right-hand side of this equation
are obtained from the conservation of probability:
\begin{eqnarray*}
 	&&\lambda_{0}=(sk^{e}_{\uparrow}+\bar{s}k^{e}_{\downarrow})\cos\theta, \quad \lambda_{1}=k^{h}_{\uparrow}\cos\theta^{h}_{\uparrow},\\
	&&\lambda_{2}=k^{h}_{\downarrow}\cos\theta^{h}_{\downarrow}, \quad \lambda_{3}=k^{e}_{\uparrow}A, \quad\lambda_{4}=k^{e}_{\downarrow}\bar{A}.
\end{eqnarray*}

\section{Results and discussion}

In this section, we present the results of numerical calculations for the
conductance of the FM/NCSC junction at zero temperature, plotted as
a function of the dimensionless quasiparticle energy
$E/\Delta_{\textrm{0}}$. We will study the effects on the tunneling conductance of three
dimensionless parameters: the Rashba spin-orbit coupling $\alpha$,
the exchange energy $I_{0}=h_{\textrm{0}}/E_{FM}$ and the Fermi
wave-vector mismatch $R$. In our calculation, we choose $\Delta_{\textrm{0}}/E_{FS}=0.01$, and
consider two cases: $Z=0$, which corresponds to a negligible potential barrier at the interface, and also
$Z=1$, corresponding to a high-transparency interface, which is
often realized in the scanning tunneling microscopy experiments.

\begin{figure}
\includegraphics[width=7cm]{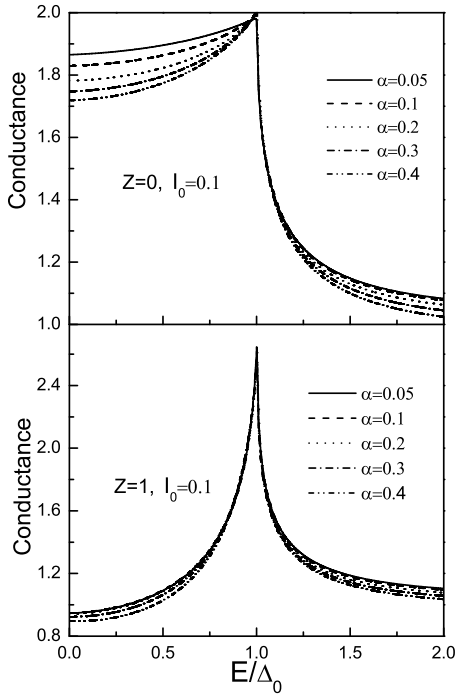}
\caption{The conductance $G(E)$ versus the dimensionless energy
$E/\Delta_{0}$ for $I_{0}=0.1$, $R=1$, and different values of the RSOC:
$\alpha=0.05$, $0.1$, $0.2$, $0.3$, and $0.4$. $Z=0$ (top panel), and
$Z=1$ (bottom panel).}
\end{figure}

We consider first the case in which there is no Fermi-surface mismatch, i.e. $E_{FM}=E_{FS}$ and $R=1$. 
Fig.~2 displays the
behavior of the tunneling conductance $G(E)$ at a fixed small
exchange energy value of $I_{0}=0.1$, for several values of $\alpha$. In the absence of the interface barrier
($Z=0$), the results are shown in the top panel. One can see clearly that the curves there are similar to the
well-known BTK results (Ref. \onlinecite{prb8225-4515}). In the BTK model, the conductance in the subgap region, $0\leq E\leq\Delta _{0}$, for the materials with
$I_{0}=\alpha=0$ is equal to 2 due to the Andreev reflection. One can see that our curves in the top panel indeed approach this
value (and are all close to 2 at the gap edge, i.e. at $E=\Delta_0$). That
the subgap conductance is slightly smaller than 2 can be attributed to the suppression of the Andreev reflection due to the different
densities of states in the spin-up and spin-down bands. The conductance at zero energy and
also far away from the gap edge monotonically decreases
with increasing the RSOC in the NCSC. This can be understood as
follows: As $\alpha$ increases, the transmitted waves with the wave
vectors $k_{1}^{e(h)}$ quickly become evanescent, since the
angle of total reflection $\theta_{c2}$ for the waves with $k_{1}^{e(h)}$
decreases as $\alpha$ increases. The eigenstates corresponding to such waves
can no longer contribute to the conductance. In the bottom
panel of Fig.~2, $Z=1$, the conductance curves display similar
behavior, but with a stronger suppression of $G(E)$ in the subgap region and a higher and sharper maximum at the gap edge $E=\Delta _{0}$.

\begin{figure}
\includegraphics[width=7cm]{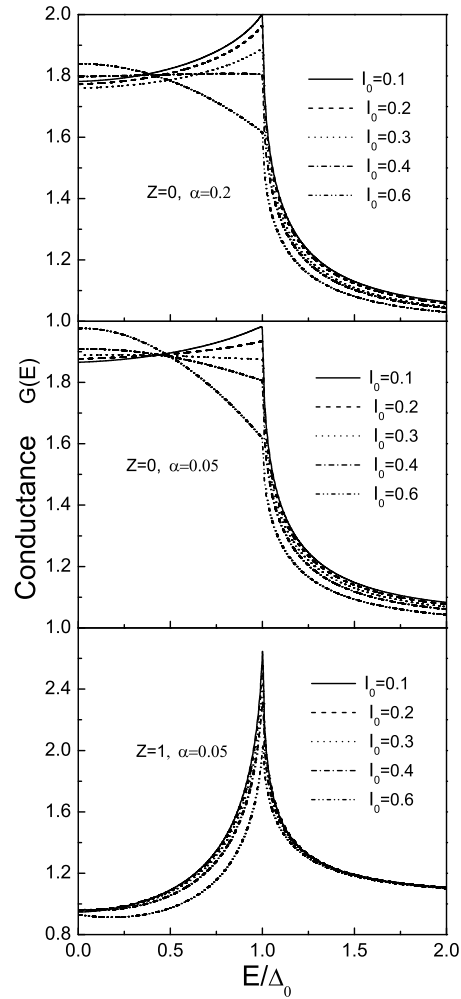}
\caption{The conductance $G(E)$ versus the dimensionless energy
$E/\Delta_{0}$ for $R=1$ and different values of the exchange energy:
$I_{0}=0.1$, $0.2$, $0.3$, $0.4$, and $0.6$. $\alpha=0.2$, $Z=0$ (top panel),
$\alpha=0.05$, $Z=0$ (middle panel), and $\alpha=0.05$, $Z=1$ (bottom
panel).}
\end{figure}

We next consider the effect of the exchange energy on the tunneling
conductance in the same situation as in the previous figure, i.e. for
$R=1$. In Fig.~3, the variation of $G(E)$ with $E/\Delta _{0}$ is
plotted for several values of $I_{0}$. In a FM/CSC junction, the conductance
monotonically decreases with increasing $I_{0}$ (Refs. \onlinecite{prl9574-1657} and \onlinecite{prb9859-9558}), because of the reduction of the Andreev
reflection, when only a small fraction of injected electrons from the majority spin band can be reflected as holes belonging to the
minority spin band. However, if the superconductor has no inversion
symmetry, the Fermi surface is split into two due to the spin-orbit coupling, thus making the conductance features more interesting. As
seen clearly from the top ($\alpha=0.2$) and middle ($\alpha=0.05$) panels in Fig.~3, in the presence of the RSOC, the exchange energy can enhance 
the Andreev reflection and therefore the subgap conductance in the region $0\leq E\leq E^*$, where $E^*\simeq\Delta _{0}/2$. This effect 
becomes more pronounced at $\alpha=0.05$, in which case
the subgap conductance at $E=0$ is monotonically enhanced for all values of $I_{0}$. The conductance can even have a maximum at $E=0$ at 
certain values of $I_{0}$ and $\alpha$. These features are quite different
from those observed in the FM/CSC junction where the peak stems from
the interplay of the FWM and the exchange
field.\cite{prb9960-6320,prb0061-1555,prb0979-014502} When the
interfacial scattering is nonzero, as shown in the bottom panel of
Fig.~3, a rather sharp conductance peak appears at the gap edge. It becomes increasingly narrow as $I_{0}$
grows, due to the suppression of the Andreev reflection. Furthermore, the
exchange energy dependence becomes weak in the region $E>\Delta_{0}$, 
and the conductance approaches its normal-state value
$G(E)=1$ (Ref. \onlinecite{prb8225-4515}) at higher excitation energies.

\begin{figure}
\includegraphics[width=7cm]{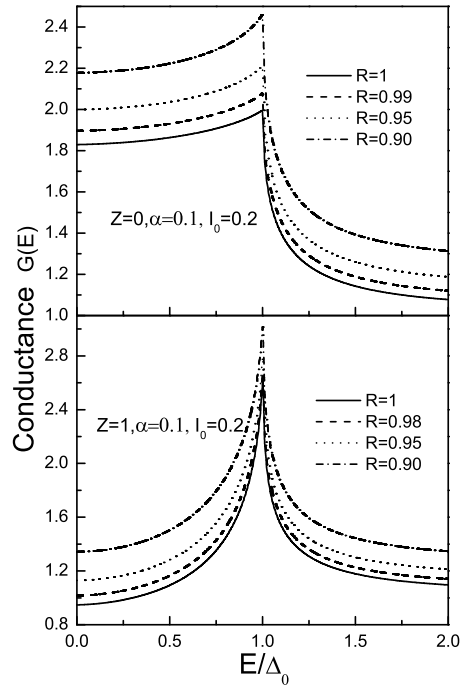}
\caption{The conductance $G(E)$ versus the dimensionless energy
$E/\Delta_{0}$ for $\alpha=0.1$, $I_{0}=0.2$, and different values of the Fermi wave-vector mismatch parameter:
$R=1$, $0.98$, $0.95$, and $0.90$. $Z=0$ (top panel) and $Z=1$ (bottom panel).}
\end{figure}

We now turn to the effects of the Fermi wave-vector mismatch, namely,
$R\neq1$, on the tunneling conductance. The difference of the
Fermi energies in the FM and NCSC regions results in some interesting features in the conductance
spectrum. In Fig.~4, which shows the results at $\alpha=0.1$,
$I_{0}=0.2$, $Z=0$ (top panel) and $Z=1$ (bottom panel), we consider
the evolution of the conductance curves for several values of the FWM.
One can easily see that the conductance is monotonically enhanced in the
whole region of excitation energies as the FWM parameter $R$ decreases (i.e. the difference between $E_{FM}$ and $E_{FS}$ increases), which is
significantly different from the case of a FM/CSC junction.\cite{prb9960-6320,prb0061-1555,prb0979-014502} This
result can be explained by the fact that in the presence of the
RSOC, a smaller $R$ will lead to the weaker ordinary scattering at
the interface, which increases the Andreev reflection. We would like to point out that in the absence of the RSOC, one cannot
obtain the monotonic increase of the conductance at all excitation energies by varying $R$ and/or $I_0$.

\section{Summary}

To summarize our results, we have investigated the tunneling conductance of the
FM/NCSC junction with the help of the extended BTK formalism. Our results show a number of features in $G(E)$ that
are qualitatively different from the previously studied cases of N/NCSC and FM/CSC junctions. 
These are caused by the interplay between the Rashba spin-orbit coupling in the noncentrosymmetric superconductor, the exchange energy in the
ferromagnet, and the Fermi wave-vector mismatch between the two
regions. 

If the Fermi energies in FM and NCSC regions are the same, then far from the gap edge the conductance is
monotonically enhanced by introducing a small RSOC, while around the
gap edge the conductance is almost independent of RSOC. In addition,
the subgap conductance can be enhanced due to the interplay
of the RSOC and the exchange energy, and can have a maximum at $E=0$
at certain values of $\alpha$ and $I_{0}$. The enhancement of the conductance is more pronounced at smaller $\alpha$, which is
attributed to the increase in the Andreev reflection by the small RSOC
dominating the decrease due to the exchange energy. These
phenomena are essentially different from those found in FM/CSC
junctions, where both the enhanced subgap conductance and its maximum
arise from the effect of the FWM at a fixed exchange energy. 

We also considered the case of different Fermi energies in the FM and NCSC
regions. The tunneling conductance is quite sensitive to the FWM and
displays a monotonic increase as the difference between the Fermi energies increases, due to the
suppressed ordinary scattering at the interface and enhanced Andreev reflection. This behavior is also essentially different from that in
FM/CSC junctions.

As for the experimental situation, while we are not aware of any work done on FM/NCSC junctions, FM/CSC junctions have been studied in Refs. \onlinecite{prl9881-3247,science98282-85,prb0367-020411}. In those works, the spin polarization of the current in the ferromagnet (the transport spin polarization)
was determined by analysing the experimental
data within the extended BTK scheme, with the total current decomposed into an unpolarized and a fully polarized
components. Our model, which includes the RSOC, can also be used in the context of
spin-polarized tunneling spectroscopy.

\section*{Acknowledgements}

This work was supported by a Discovery Grant from the Natural
Sciences and Engineering Research Council of Canada.


\begin{thebibliography}{99}

\bibitem{prl9472-1526} C. R. Hu, Phys. Rev. Lett. \textbf{72}, 1526 (1994).

\bibitem{prb9653-2667} S. Kashiwaya, Y. Tanaka, M. Koyanagi, and K. Kajimura, Phys. Rev. B \textbf{53}, 2667 (1996).

\bibitem{prb9756-13746} J. W. Ekin, Y. Xu, S. Mao, T. Venkatesan, D.W. Face, M. Eddy,
                        and S. A. Wolf, Phys. Rev. B \textbf{56}, 13746 (1997).

\bibitem{prl9881-3247} S.K. Upadhyay, A. Palanisami, R. N. Louie, and R. A. Buhrman,
                       Phys. Rev. Lett. \textbf{81}, 3247 (1998).

\bibitem{science98282-85} R. J. Soulen, Jr., J. M. Byers, M. S. Osofsky, B. Nadgorny, T. Ambrose,
                          S. F. Cheng, P. R. Broussard, C. T. Tanaka, J. Nowak, J. S. Moodera, A. Barry,
                          and J. M. D. Coey, Science \textbf{282}, 85 (1998).

\bibitem{prb0367-020411} P. Raychaudhuri, A. P. Mackenzie, J. W. Reiner, and M. R. Beasley, Phys. Rev. B \textbf{67},
                         020411 (2003).

\bibitem{prl9574-1657} M. J. M. de Jong and C. W. J. Beenakker, Phys. Rev. Lett. \textbf{74}, 1657 (1995).

\bibitem{prb9859-9558} J.-X. Zhu, B. Friedman, and C.S. Ting, Phys. Rev. B \textbf{59}, 9558 (1998).

\bibitem{prb0367-174501} T. Hirai, N. Yoshida, Y. Asano, J. Inoue, and S. Kashiwaya, Phys. Rev. B. \textbf{67}, 174501 (2003).

\bibitem{prb0775-134509} J. Linder and A. Sudb{\o}, Phys. Rev. B \textbf{75}, 134509 (2007).

\bibitem{prb9960-6320} I. Z\u{u}ti\'{c} and O. T. Valls, Phys. Rev. B \textbf{60}, 6320 (1999).

\bibitem{prb0061-1555} I. Z\u{u}ti\'{c} and O. T. Valls, Phys. Rev. B \textbf{61}, 1555 (2000).

\bibitem{prb0265-174508} G. Sun, D. Y. Xing, J. M. Dong, and M. Liu, Phys. Rev. B \textbf{65}, 174508 (2002)

\bibitem{jap08104-063903} Y. C. Tao and J. G. Hu, J. Appl. Phys. \textbf{104}, 063903 (2008).

\bibitem{prl0492-027003} E. Bauer, G. Hilscher, H. Michor, Ch. Paul, E. W. Scheidt, A. Gribanov, Yu. Seropegin, H.
                         No\"{e}l, M. Sigrist, and P. Rogl, Phys. Rev. Lett. \textbf{92}, 027003 (2004).

\bibitem{prl0187-037004} L. P. Gor'kov and E. I. Rashba, Phys. Rev. Lett. \textbf{87}, 037004 (2001).

\bibitem{prb0265-144508} S. K. Yip, Phys. Rev. B \textbf{65}, 144508 (2001).

\bibitem{prl0594-027004} K. V. Samokhin, Phys. Rev. Lett. \textbf{94}, 027004 (2005); Phys. Rev. B. \textbf{76}, 094516 (2007).

\bibitem{prl9575-2004} V. M. Edelstein, Phys. Rev. Lett. \textbf{75}, 2004 (1995).

\bibitem{prb0572-024515} S. Fujimoto, Phys. Rev. B \textbf{72}, 024515 (2005).

\bibitem{pc03387-13} D. F. Agterberg, Physica C \textbf{387}, 13 (2003).

\bibitem{prb0470-104521} K. V. Samokhin, Phys. Rev. B \textbf{70}, 104521 (2004).

\bibitem{prl0594-137002} R. P. Kaur, D. F. Agterberg, and M. Sigrist, Phys. Rev. Lett. \textbf{94}, 137002 (2005).

\bibitem{prb0572-220504} T. Yokoyama, Y. Tanaka, and J. Inoue, Phys. Rev. B \textbf{72}, 220504 (2005).

\bibitem{prb0776-012501} C. Iniotakis, N. Hayashi, Y. Sawa, T. Yokoyama, U. May, Y. Tanaka, and
                         M. Sigrist, Phys. Rev. B \textbf{76}, 012501 (2007).

\bibitem{prb0776-054511} J. Linder and A. Sudb{\o}, Phys. Rev. B \textbf{76}, 054511 (2007).

\bibitem{prb8225-4515} G. E. Blonder, M. Tinkham, and T. M. Klapwijk, Phys. Rev. B \textbf{25}, 4515 (1982).

\bibitem{prb0979-014502} P. H. Barsic and O. T. Valls, Phys. Rev. B \textbf{79}, 014502 (2009).

\bibitem{prb0164-121202} L. W. Molenkamp, G. Schmidt, and G.E.W Bauer, Phys. Rev. B \textbf{64}, 121202 (2001).

\end{thebibliography}
\end{document}